\documentclass[fleqn,usenatbib]{mnras}

\usepackage{newtxtext,newtxmath}
 
\usepackage[T1]{fontenc}
\DeclareUnicodeCharacter{2061}{}
\pdfoutput=1
\DeclareRobustCommand{\VAN}[3]{#2}
\let\VANthebibliography\thebibliography
\def\thebibliography{\DeclareRobustCommand{\VAN}[3]{##3}\VANthebibliography}

\usepackage{graphicx}	
\usepackage{amsmath}	
\usepackage{amssymb}	

\title[Characterizing RFSoC for Radio Astronomy Receivers]{Characterising the Performance of High-Speed Data Converters for RFSoC-based Radio Astronomy Receivers}

\author[C. Liu et al.]{
Chao Liu,$^{1}$\thanks{E-mail: chao.liu@physics.ox.ac.uk}
Michael E. Jones,$^{1}$
and Angela C. Taylor$^{1}$
\\
$^{1}$Department of Physics, University Of Oxford, Oxford, OX1 3RH, United Kingdom
}

\date{Accepted XXX. Received YYY; in original form ZZZ}

\pubyear{2020}

\begin{document}
\label{firstpage}
\pagerange{\pageref{firstpage}--\pageref{lastpage}}
\maketitle

\begin{abstract}
RF system-on-chip (RFSoC) devices provide the potential for implementing a complete radio astronomy receiver on a single board, but performance of the integrated analogue-to-digital converters is critical. We have evaluated the performance of the data converters in the Xilinx ZU28DR RFSoC, which are 12-bit, 8-fold interleaved converters with a maximum sample speed of 4.096 Giga-sample per second (GSPS). We measured the spurious-free dynamic range (SFDR), signal-to-noise and distortion (SINAD), effective number of bits (ENOB), intermodulation distortion (IMD) and cross-talk between adjacent channels over the bandwidth of 2.048 GHz. We both captured data for off-line analysis with floating-point arithmetic, and implemented a real-time integer arithmetic spectrometer on the RFSoC. The performance of the ADCs is sufficient for radio astronomy applications and close to the vendor specifications in most of the scenarios. We have carried out spectral integrations of up to 100 s and stability tests over tens of hours and find thermal noise-limited performance over these timescales.
\end{abstract}

\begin{keywords}
Instrumentation:miscellaneous -- techniques:miscellaneous
\end{keywords}



\section{Introduction}

Radio astronomy receivers have been developing along a steady trend towards earlier digitization, greater integration, and more use of commercial hardware. Whereas once a radio astronomy backend system might have used custom-designed analogue-to-digital converters (ADCs)  and application-specific integrated circuits (ASICs) \citep[e.g.][]{2007A&A...462..801E}, in recent systems ASICs have largely been replaced by field-programmable gate arrays (FPGAs), which now provide sufficient processing power for real-time channelization and other digital signal processing (DSP) operations over several GHz of bandwidth \citep[e.g.][]{2016mks..confE...1J}. 

\textbf{}Early-generation FPGA systems required a CPU chip on the board to provide local control of the FPGA, but now `system-on-chip' (SoC) devices integrate CPU cores on the same device as the FPGA, providing software control closely coupled to the programmable logic system. High-speed data converters have however still typically been separate devices, and despite the establishment of interface protocols such as JESD-204B\footnote{http://www.jedec.org}, interfacing them to FPGAs has been a significant design effort and has absorbed a great deal of effort in the community. However, driven by the needs of the telecommunications industry, the FPGA manufacturer Xilinx has now integrated high-speed data converters in to the same chip as the FPGA and CPU cores, producing a fully-digital `RF system-on-chip' (RFSoC). Previous RFSoC devices have integrated RF components such as low-noise amplifiers, mixers, filters etc, along with programmable logic and microcontrollers, but the very high sample rate and DSP capability of the new devices renders the RF components redundant, as all signal functions can now be implemented digitally. These devices now combine all the major functions of a receiver, for radio astronomy or other applications, in a single package -- digitization, real-time signal processing, software control, and high-speed interfacing to the downstream system. 

We have several specific applications of the RFSoC-based system in mind. These include a project to upgrade the analogue backend of the ultra-wideband Redshift Search Receiver (RSR) \citep{erickson2007} on the Large Millimetre Telescope (LMT) to a digital one. The RSR covers 74 –110.5 GHz and it has four receiver channels, with a total IF bandwidth of 146 GHz. The entire band needs to be spectrally channelized simultaneously. We also plan to upgrade the C-Band All-Sky Survey (C-BASS) South receiver \citep{jones2018c}. The current C-BASS South receiver uses two ROACH boards to provide 128 frequency channels across a 4.5 -- 5.5 GHz bandwidth. With a suitable wide-band digital backend we plan to implement a 7 -- 15 GHz receiver to carry out an X-Band All-Sky Survey (X-BASS) \citep{jaz}. With appropriate downconversion hardware, both these projects, and others, could use the same RFSoC platform and firmware for their backends. However, before embarking on these implementations we need to determine if the Xilinx RFSoC data converters have adequate performance for radio astronomy applications. In this paper we describe the evaluation we have carried out of the data converters. In Section \ref{hardware} we describe the hardware systems used for the tests. In Section \ref{sec:ADCperformance} we describe tests of the ADC performance made by capturing data and processing it offline with floating-point arithmetic, while in Section \ref{sec:spectrometer} we describe tests using real-time integer processing on the FPGA. We conclude in Section \ref{sec:conclusions}.

\section{Hardware platform}\label{hardware}

The Zynq UltraScale RFSoC series of devices from Xilinx caught the attention of the radio astronomy community when it was first announced in 2017. Three generations of devices have now been announced, with sampling speeds up to 5.0 Giga-samples per second (GSPS). The device used here for evaluation is one of the RFSoC first-generation devices from Xilinx, the ZU28DR, which has eight 12-bit RF ADCs with sampling frequency up to 4.096 GHz integrated with an FPGA fabric and two processors in a single device. The device offers a large amount of logic cells, DSP slices and memory, with sixteen 33Gb/s transceivers, two 100G Ethernet interfaces, a Quad-core Arm Cortex-A53 processor and a dual-core Arm Cortex-R5 processor. The FPGA fabric and associated transceiver hardware are referred to as the programmable logic (PL), while the processor cores are referred to as the processing system (PS). The combination of these key components makes this RFSoC a suitable candidate platform for many radio astronomy receivers.  The components of the RFSoC could form a complete radio astronomy spectrometer for single-dish applications, or a data acquisition, pre-processing and packetization system for interferometers: the RF ADCs can digitize the IF signals, the programmable logic and DSP resources provide data processing, the transceivers provide high-throughput data transfer, and the processors provide for system management and configuration.  The high integration level of RFSoC can significantly simplify the design and implementation processes and reduce the hardware costs of radio astronomy backend development and scaling up.

The development of RFSoC family is largely driven by the increasing demand in wideband communication, so test data from the manufacturer is only available at some of the frequency bands of interest for communication systems, rather than the entire bandwidth. For wide-band radio astronomy applications, it is essential to have performance evaluation over the entire bandwidth of the ADCs. Therefore, the evaluation summarised in this paper covers the full bandwidth of the ADCs. 

The ADCs integrated in these RFSoC parts have an interleaved structure. Each of the 4.096 GHz ADC has eight 512 MHz sub-ADC slices \citep{farley2017programmable}. That raises concerns about interleaving spurs, and the overall performance. This paper summarises a comprehensive ADC evaluation with most of the commonly studied parameters. We have evaluated the ADCs using floating-point, off-line post-processing, to isolate the performance of the ADCs from any concerns about fixed-point arithmetic in the FPGA,  but the evaluation has also been extended using a single-channel spectrometer implemented in the FPGA operating in real-time. The ADC has demonstrated satisfactory performance for radio astronomy receivers even with over 100s of integration time.

The evaluation board ZCU111, produced by Xilinx, was used as the evaluation platform. The ZCU111 board is populated with the XCZU28DR-2FFVG1517E RFSoC device. There are four RF ADC tiles, numbered from 224 to 227, in each device. Each of the tiles consists of two ADCs, ADC01 and ADC23, and both of them sample up to 4.096 GHz. The full-scale input of the ADC is 1 V peak-to-peak, corresponding to a CW input power of $+4$ dBm. The XCZU28DR device has sixteen 32.75 Gb/s transceivers on the programmable logic. The transceivers are grouped in fours, and each group is called a quad. One of quads has been connected to four SFP28 connectors and the other three are connected to an FMC+ connector on the board. A high-speed data interface with 100 Gbps or above can be realized by using the SPF28 connectors, while a FMC+ module with high-speed connectors or any other connectors connected to those transceivers. 

The ZCU111 board is not ideal for use in a production radio astronomy system. To facilitate evaluation of different sampler and communications configurations, the RF and transceiver connections are brought out to a variety of different connector and interface types. However, other vendors produce boards with more consistent interfaces which are better suited to production systems, with, for example the same type of baluns on each RF ADC input. This does not prevent us from investigating the fundamental performance of the RFSoC device using the ZCU111. 

\begin{figure}
\includegraphics[width=\columnwidth]{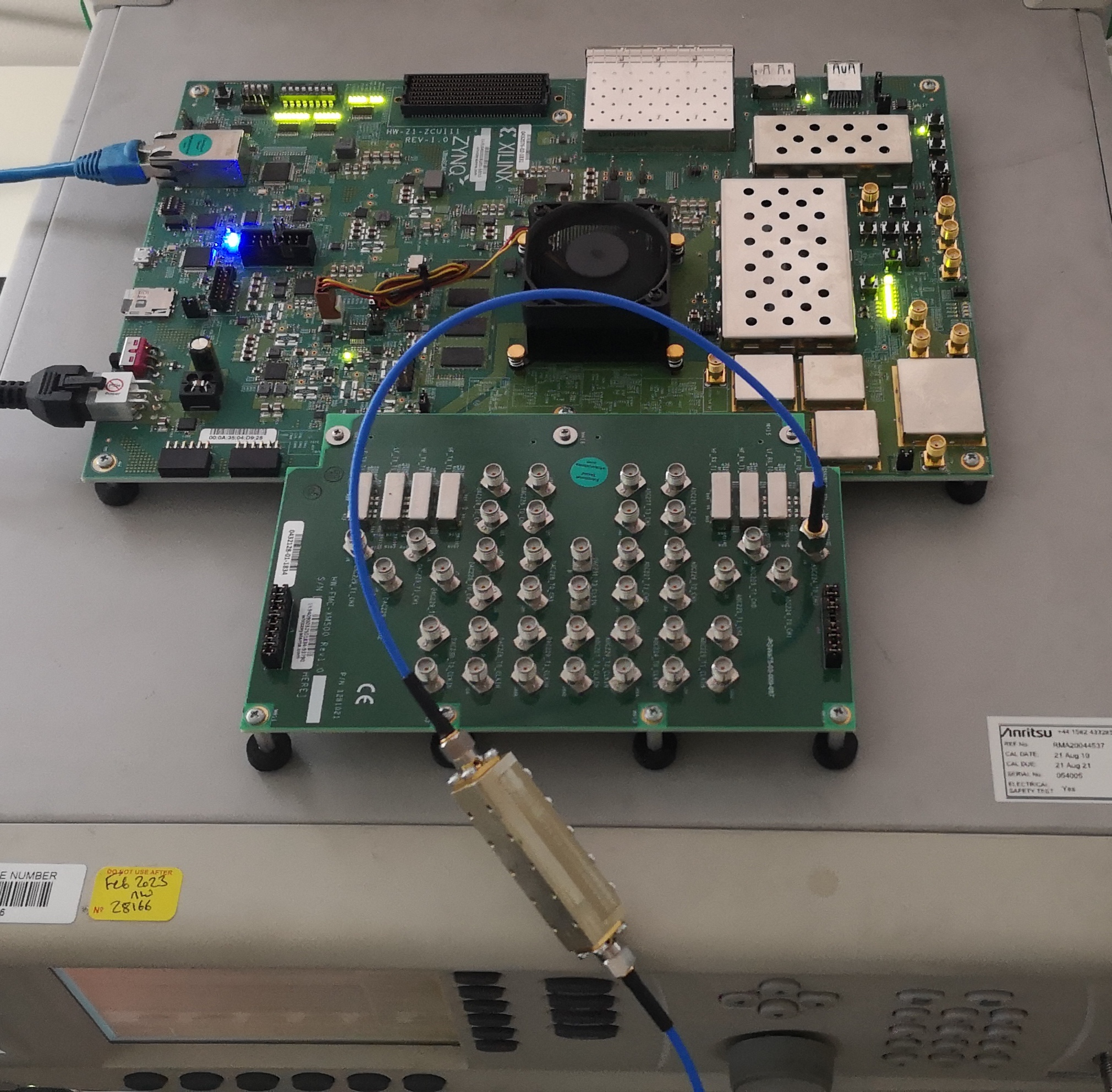}
\caption{Photograph of the ZCU111 evaluation board with one ADC channel connected to a signal generator.}
\label{aba:fig1}
\end{figure}

\section{Data converter performance evaluation}\label{sec:ADCperformance}

\subsection{Testing setup}\label{sec:setup}

On the ZCU111 evaluation board, the differential input pairs of the RFSoC ADCs are routed to a daughter board connected via an FMC+ connector.  On the daughter board, two of the ADC input differential pairs are converted to single-ended signals using an RF transformer, a TCM2-33WX+, which has a bandwidth of 10 MHz to 3 GHz. The typical insertion loss of TCM2-33WX+ is 1.5 dB over the bandwidth.

The RF test signal to be injected to the ADCs was generated using an Anritsu MG3692B signal generator. An anti-aliasing low pass filter (LPF) with 1.6 GHz cut-off frequency, designed in-house, was connected in series with the test signal before it was connected to the SMA port on the daughter board. This LPF was used in all the tests summarised in this section, except Section \ref{sec:noise_floor}.

The ADC01 in tile 224 was enabled and configured to sample at 4.096 GSPS. Data capturing firmware/software was implemented on the RFSoC. The RFSoC RF Data Converter IP, instantiated via the Xilinx Vivado toolflow, has an AXI-streaming data bus 128 bits wide with 512 MHz data rate. A block of ADC samples is first written to a FIFO to reduce the data rate to 256 MHz and hence ease the timing restrictions. Then the data are written to DDR4 memory, which is accessible to the processing system. The data stored in DDR4 are packetized and sent to the host computer via Ethernet. The software application was implemented in the FreeRTOS operating system with the support of DMA driver, I$^2$C driver, RFdc driver and Lightweight socket API. The software application configures the off-chip PLL to generate the clocks for the data converters through an I$^2$C interface, sets and checks the mode register of the data converters, manages the data movement between programmable logic and processing system, and packetizes and sends ADC samples as UDP packets to the host computer. 

On the host PC, the UDP packets are received and saved by a Matlab script. The data received are interpreted by routines realized as Matlab scripts, and include Fast Fourier Transform (FFT), power accumulation and division to get the parameters and plots desired.

\subsection{Noise floor and calibration}\label{sec:noise_floor}

Before proceeding with signal tests of the ADC, we checked the noise level of the converter with zero input power. The input of ADC01 on tile 224 was terminated by a 50 $\Omega$ terminator. The ADC was configured to sample at 4.096 GSPS and ADC samples were captured as blocks of 65,536 samples. The sample blocks were divided in to 16 sub-blocks with 4,096 samples each. The FFT was computed for each sub-block and the spectra obtained were averaged. The goal was to have a clear definition of the noise floor of the ADCs. 

\begin{figure}

\includegraphics[width=\columnwidth] {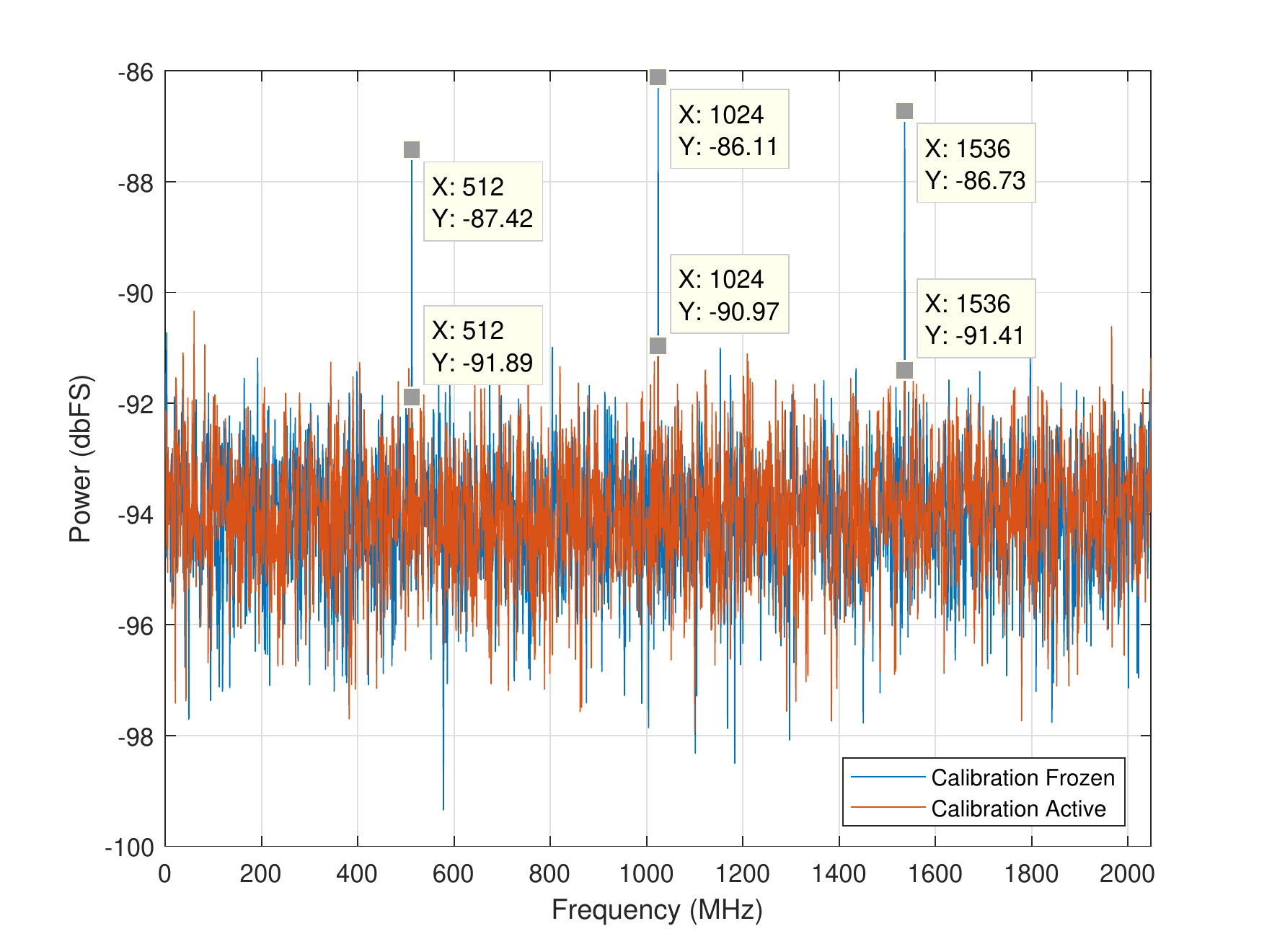}

\caption{Noise floor with ADC calibration frozen and active. Note the interleaving spurs at multiples of 512 MHz when the calibration is not active.}
\label{aba:fig2}
\end{figure}

The ADCs have an interleaved structure, with each sub-ADC slice being sampled at one-eighth of the overall ADC sample rate, i.e. 512 MHz. Avoiding spurious responses at this frequency requires calibration to align the sampling offsets for each sub-ADC slice before starting normal operation. The calibration routine is implemented as part of the RFSoC RF Data Converter IP core provided by Xilinx, and the routine can be enabled and disabled by the configurations of the IP. In this test, the calibration was enabled and disabled to explore the effect of the routine on the noise floor. The spectra with calibration frozen and calibration active are shown in Figure \ref{aba:fig2}. As the figure shows, the interleaving spurs at 512 MHz and its multiples are obvious when the calibration was frozen. The interleaved spurs are reduced to the noise floor level by activating the calibration. Therefore, the calibration was kept active in the rest of the tests carried out in this paper and should always be enabled in applications of the RFSoC. 

\subsection{Spurious-free dynamic range (SFDR)}\label{sec:SFDR}

Spurious-free dynamic range (SFDR) is the ratio of the power in the fundamental tone to the power of the worst spur in the frequency range under investigation. SFDR is a critical specification for radio astronomy receivers as it determines the smallest signal that can be observed over the spectrum of interest. 

\begin{figure}
\includegraphics[width=\columnwidth] {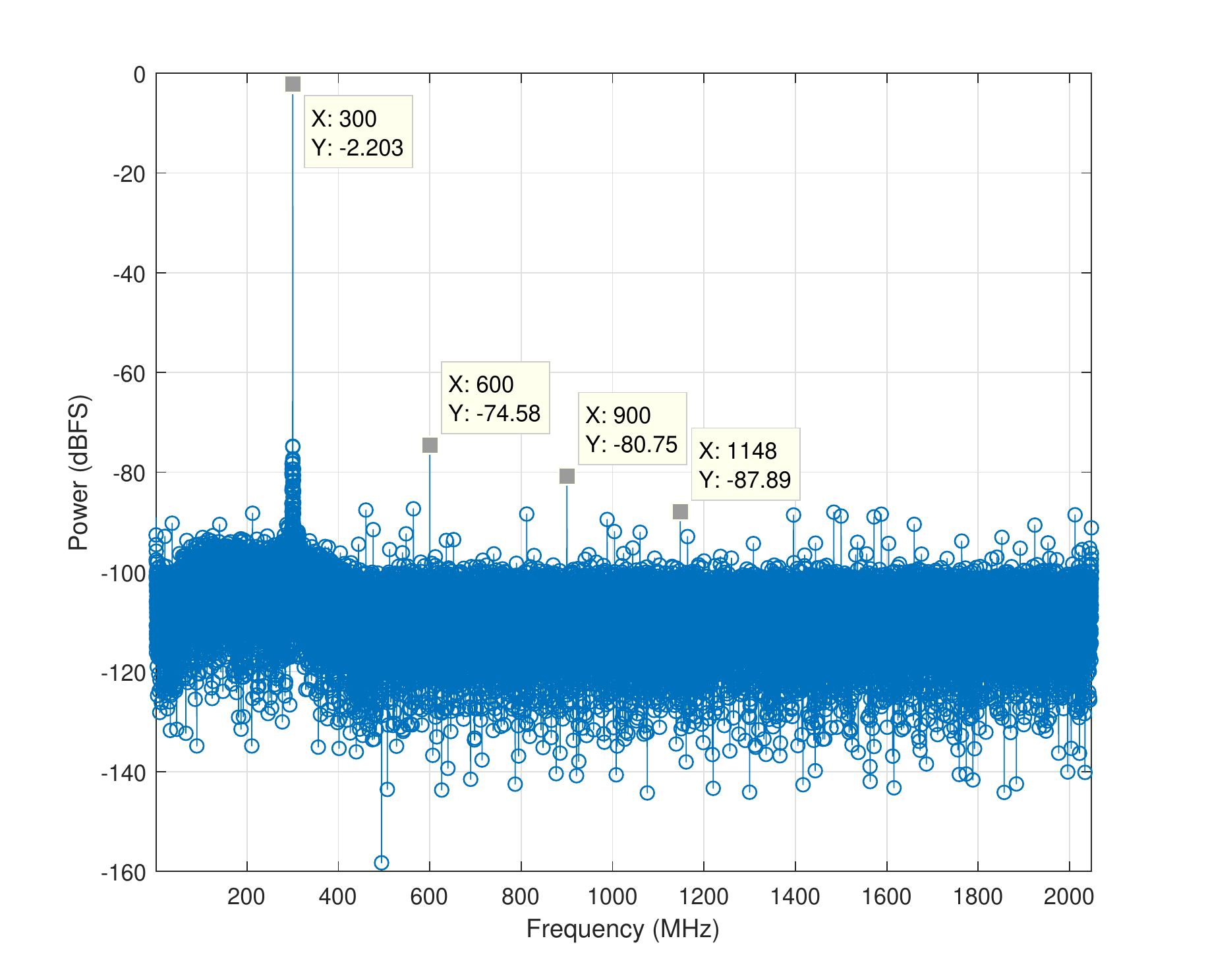}
\caption{Typical spectrum used for measurement of SFDR, with tone frequency set to 300 MHz.}
\label{fig:spec}
\end{figure}

In this test, 65,536 ADC samples were captured as a block and an FFT performed on the entire block, so the frequency resolution of the spectrum obtained over the 2.048-GHz bandwidth is 62.5~kHz. All the computation performed in Matlab uses floating-point numbers in double precision. The same FFT routine is used for all the tests in this section, except Section \ref{sec:noise_floor}. The tones generated by the signal generator start from 300 MHz and increase in 300~MHz steps up to  1.5~GHz. The output power of the tones was set to $+5$~dBm at the signal generator. However, due to the insertion loss of the RF transformer and loss along the tracks, the power in the tones captured by the ADC is $-2$~dBFS at 300~MHz and declines as frequency goes up. Figure \ref{fig:spec} shows a typical spectrum with the tone frequency set to 300~MHz. As the figure shows, the harmonics of the fundamental are significantly higher than any other spurs. The harmonics dominate the noise, so the SFDRs are largely the differences between the power of the fundamental tone and its harmonics.  

\begin{figure}
\includegraphics[width=\columnwidth] {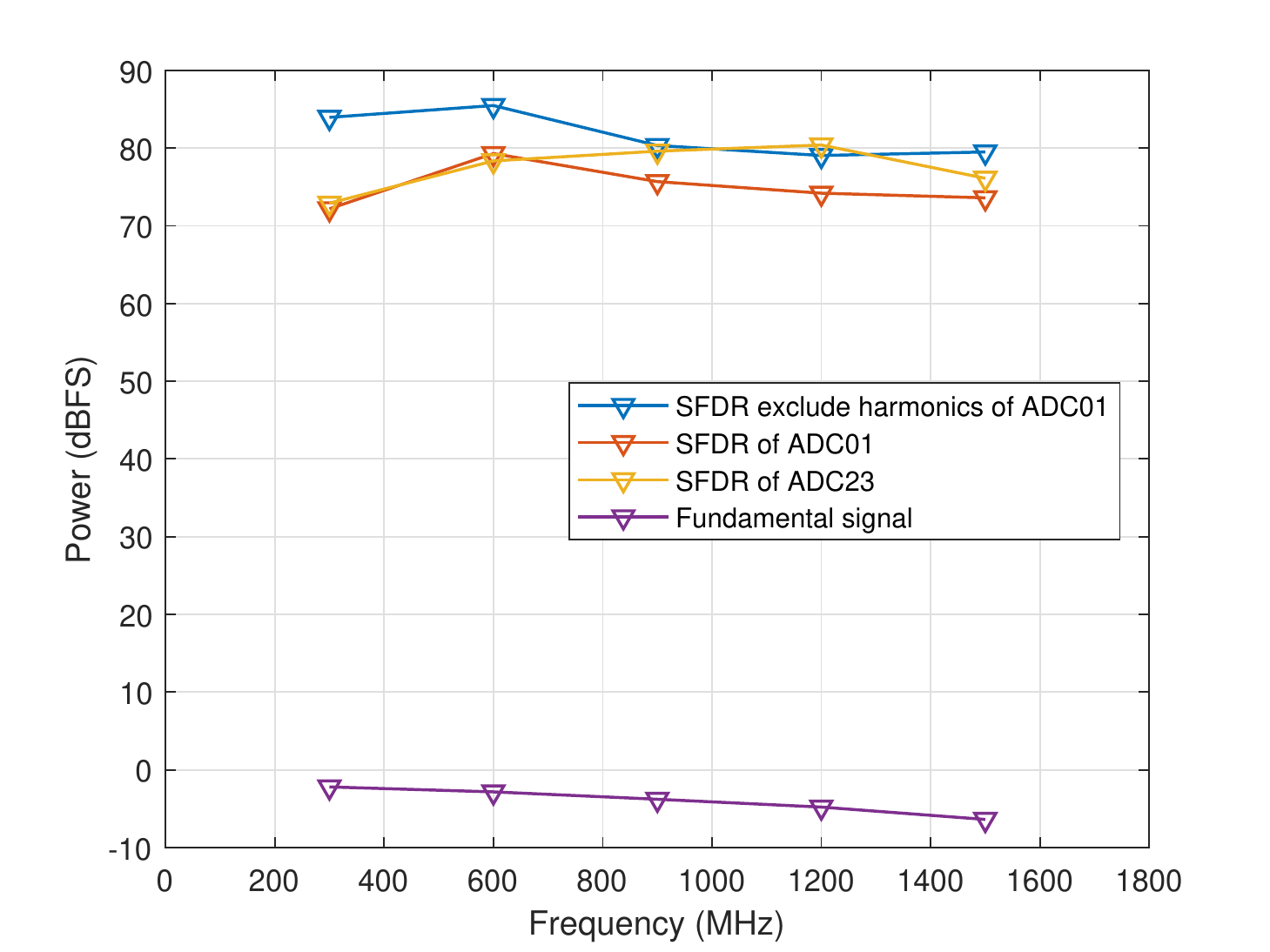}
\caption{Fundamental tone power level from the signal generator, measured SFDRs from two ADC tiles, and SFDR excluding harmonics of the fundamental. }
\label{fig:SFDR}
\end{figure}

As Figure \ref{fig:SFDR} shows, the SFDR is around 72 to 80~dB across the effective frequency range of the ADC. The SFDR is a direct reflection of the effective dynamic range of the ADC, and a dynamic range at this level is adequate for the majority of radio astronomy applications.

For a wide-band radio astronomy receiver, the ADCs will be used to digitize different IF segments of the RF signal after downconversion, so consistency of performance among the ADCs on the RFSoC is crucial. The same test has been performed for ADC23 on tile 224, which is adjacent to ADC01. As Figure \ref{fig:SFDR} shows, the SFDR of ADC23 is marginally lower than ADC01 at 600 MHz, and higher at all other frequencies, with both tiles having an SFDR exceeding 70~dB. Since the performance of these ADCs are similar, the IF segments digitized by them could be combined to a wide-band instrument with consistent performance in the entire bandwidth it covers. The rest of the ADCs in the RFSoC have different input circuits on the evaluation daughter board, so the comparison could not be extended to ADCs in other tiles. However, the comparison will be made between all the ADCs  using a board with identical input circuits in a system we are currently developing.  

The SFDRs listed in the Xilinx data sheet exclude the second and third harmonics of the test frequency. The blue trace in Figure \ref{fig:SFDR} shows the SFDR excluding those harmonics, which is around 80 to 85~dB. The results are close to the specifications of 86~dB at 240 MHz and 83 dB at 1.9 GHz listed in the Xilinx data sheet \citep{xilinxacdc}. The results in the data sheet are measured with an input power of $-1$~dBFS. The lower SFDR excluding harmonics can be partially attributed to the lower input power level, especially at higher frequency end. The power levels of the tones generated were not tuned frequency by frequency to achieve  $-1$~dBFS level at the ADC input. However, the constant input power level is useful to show the effect of the input circuit losses on the system performance. The test results might also be improved by applying a narrow band pass filter at each of the individual test tones, which can damp the harmonics as well as other noise signals generated by the signal generator. However, even without these precautions, the measured SFDR is extremely high.

When the SFDR excluding the harmonics is measured, we find that the interleaving spurs at multiples of 512 MHz in some of the tests are higher than the noise floor. However, the interleaved spurs measured  are below the levels specified in the Xilinx data sheet in the frequency range we interested in, which are $-81$~dBc at 240~MHz and $-80$~dBc at 1.9~GHz.

\subsection{ADC SFDR using the RFSoC DACs}

The RFSoC is designed for software-defined radio applications with full transmit and receive capability. Hence there are also eight digital-to-analogue converters (DACs) with sample rates up to 6.554~GHz integrated in the XCZU28DR. In the context of astronomy receivers these DACs can be used to generate test signals, and also can be used in receiver systems such as those needed for Kinetic Inductance Detectors (KIDs) \citep{van2016multiplexed}, and as part of the calibration process for some radio astronomy receivers. The DACs were configured to generate single-frequency tones and the signals looped back to the ADCs to characterize both of the data converters. The DACs have two programmable output levels, 20~mA (corresponding to $+1$~dBm) and 32 mA ($+5$~dBm). We used an output power level of  $+1$~dBm and a  sample rate of 4.096 GSPS. The same set of frequency tones were  used as in Section \ref{sec:SFDR}, and the same anti-aliasing LPF was inserted before the signal was connected to the single-ended input of the ADC. 

\begin{figure}
\includegraphics[width=\columnwidth] {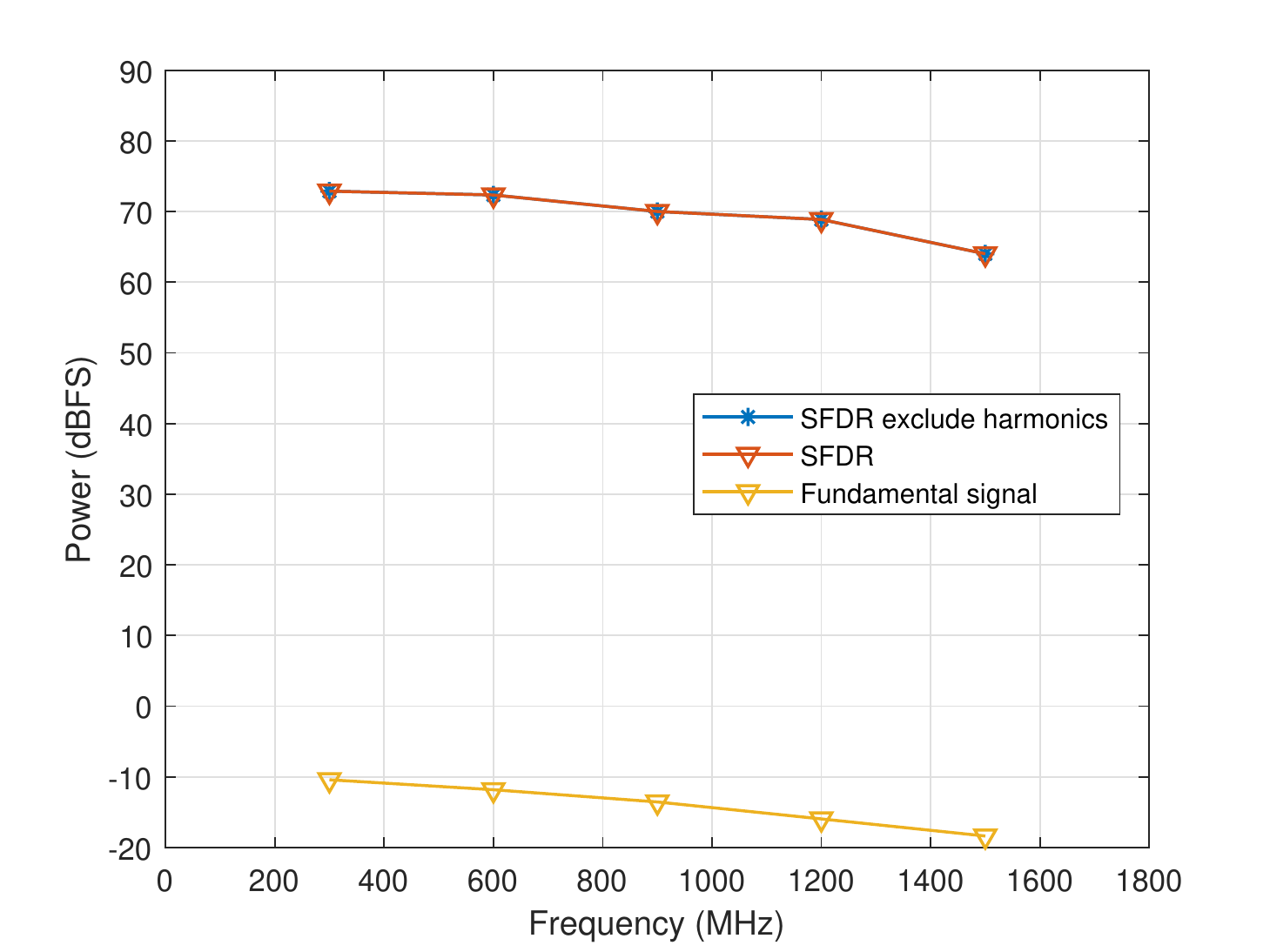}
\caption{Fundamental tone power level and SFDRs measured with signals generated by the DAC, configured to 25 mA output current.}
\label{aba:fig4}
\end{figure}

\begin{figure}
\includegraphics[width=\columnwidth] {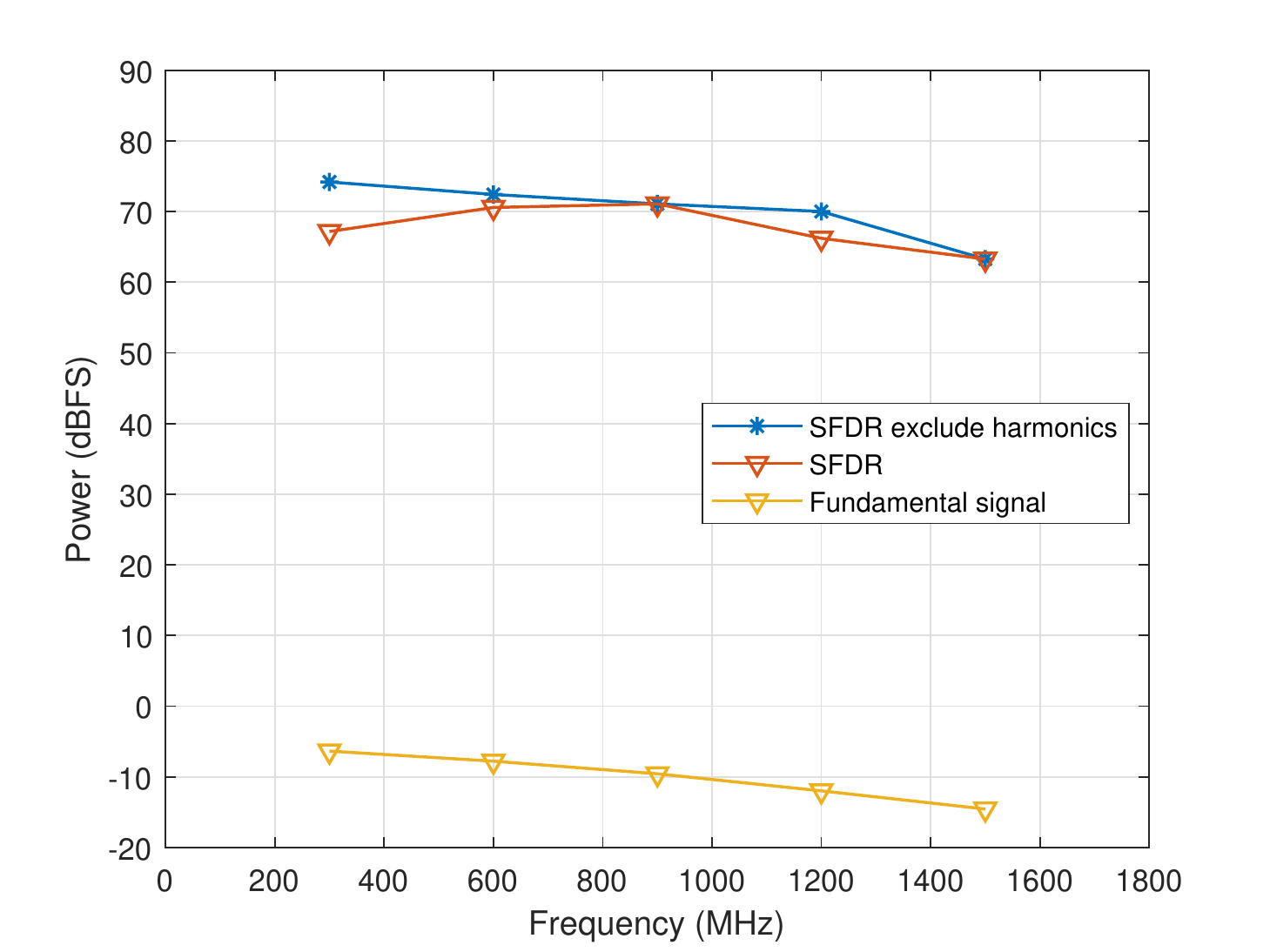}
\caption{Fundamental tone power level and SFDRs measured with DAC configured to 35 mA output current.}
\label{aba:fig5}
\end{figure}

Figures \ref{aba:fig4} and \ref{aba:fig5} show the evaluation results for the two output current levels of the DAC respectively. As Figure \ref{aba:fig4} shows, the power of the fundamental component varies between $-10$ to $-20$~dBFS across the frequency range measured when the DAC output current is set to be 20~mA. There are two reasons why this is significantly lower than the power level obtainable with an external signal generator. Firstly, the power level of the signal generated by the DAC is 4~dB lower than the signal generator. Secondly, similarly to the ADC input circuit, the same RF transformer has been used to convert the differential output to single-ended one, so the insertion loss due to the RF transformers and the signal traces is doubled.  The SFDR achieved is in the range of 65 to 72 dB, which is significantly lower than the specified SFDR achieved when using the signal generator to generate the test signal. There is no direct test result in the Xilinx data sheet at this frequency and at this input power level for the ADC. However, the Xilinx data sheet specifies the SFDR values for the second Nyquist zone of ADC input at $-10$ and $-20$~dBFS of 78 and 70~dB respectively. This shows that the result in Figure \ref{aba:fig4} is acceptable. 

When the current of the DAC is set to 35 mA, the power of the tones generated improves. However, the SFDR calculated for the ADC was not improved by the increase in input power level, as shown in Figure \ref{aba:fig5}. This is due to the high level of spurs generated by the excess current. The results provide guidance for selecting from the two modes of DAC operation based on application scenarios. The test results also give the real output level of the DACs at 25 and 35 mA output current levels, and amplification circuitry can be inserted accordingly.


\subsection{Signal-to-noise and distortion (SINAD) and effective-number-of-bits (ENOBs)}

Signal-to-noise and distortion (SINAD) is a comprehensive indicator of the overall dynamic range of an ADC, as it takes both noise and distortions into account. The SINAD is the ratio of rms of the fundamental tone and the mean value of root sum square (RSS) of all other components in the spectrum. The effective number of bits (ENOB) can be calculated from  

\begin{equation}
ENOB  = \frac{SINAD-1.76 dB}{6.02},\label{aba:eq1}
\end{equation}

\noindent using the measured noise plus distortion level to give a direct expression of the effective quantization noise performance of the ADC. A Matlab script was to calculate ENOB based on Equation \ref{aba:eq1} and other parameter definitions in \cite{kester2009understand}.

\begin{figure}
\includegraphics[width=\columnwidth] {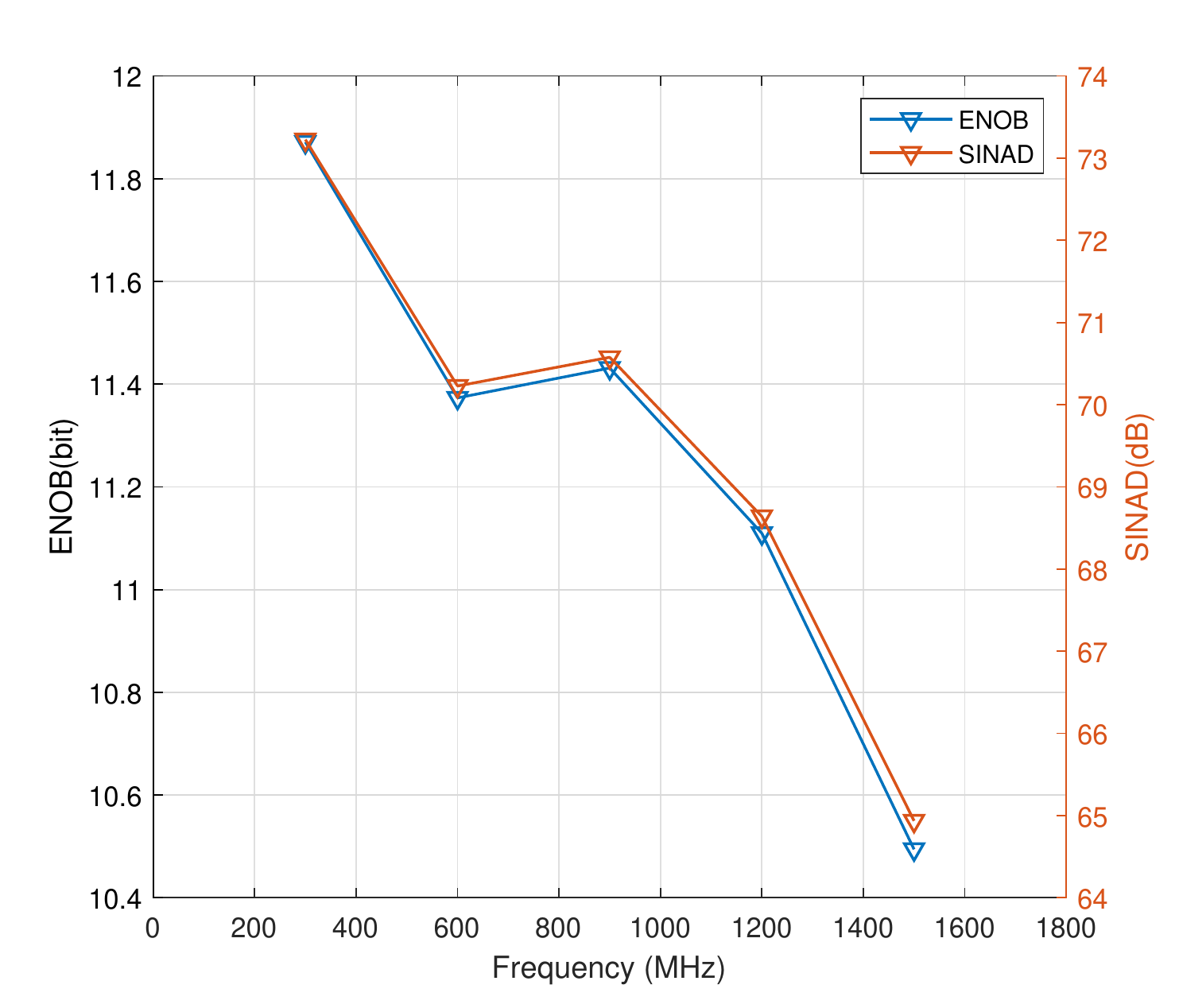}
\caption{SINAD and ENOB of ADC, with tones generated by signal generator from 300 to 1500~MHz.}
\label{aba:fig6}
\end{figure}

The ADC samples used in Section \ref{sec:SFDR} to calculate SFDR were used to compute the ENOB and SINAD shown in Figure \ref{aba:fig6}. The ENOB, shown on the left axis of the plot, decreases from 11.87 to 10.49 bits as the input frequency increases. The ENOB at 300 MHz is very close the ideal 12 bits resolution of the ADC, and about 1.5 bit worse at 1.5 GHz. This can be partially attributed to the decreasing power level of the fundamental tone as frequency increases, and this is also reflected on the SINAD shown in Figure \ref{aba:fig6}. There are no test results for SINAD and ENOB listed in the Xilinx data sheet, so we cannot compare directly to the manufacturer's data. However, the loss of effective bits is comparable to other high-speed data converters, and provides more than enough dynamic range for most radio astronomy applications. 

\subsection{Intermodulation distortion (IMD)}

The ADC linearity performance is critical for a sensitive radio astronomy spectrometer or interferometer. The linearity of ADCs is commonly measured by intermodulation distortions (IMDs) rather than the third-order intercept point or 1-dB compression point, which are more useful concepts for fully analogue devices such as amplifiers \citep{kester2009intermodulation}. IMDs are measured by applying two closely-spaced tones of equal amplitude $\cos(\omega_1 t)$ and $\cos(\omega_1 t)$, and finding the maximum amplitude of components corresponding to either $(\cos(\omega_1 t) + \cos(\omega_2 t))^2$ (IMD2) or $(\cos(\omega_1 t) + \cos(\omega_2 t))^3$ (IMD3). The two tones for testing were generated by the DAC integrated in the RFSoC at its full output power, looped back to the ADC input via the anti-aliasing filter. The centre frequencies, frequency gaps and power levels of the two-tone signals used for testing, together with the IMD2 and IMD3 obtained from the FFT of captured ADC samples are summarised in Table \ref{tab:tbl1}. The listed max IMD3 at -7~dBFS input in this frequency range in the Xilinx data sheet is -68~dBc, which is close to the IMD3 values obtained in this case. The IMD2s are the dominant noise spurs and therefore can be regarded as the multi-tone SFDR, at around $-60$~dBc. This sets the effective dynamic range in the presence of multiple strong, narrow-band signals such as RFI sources.   

\begin{table}
	\centering
	\caption{Intermodulation test setup and results summary.}
	\label{tab:tbl1}
	\begin{tabular}{lccr} 
		\hline
		Centre Frequency (MHz)       & 1024              & 512\\
		\hline
		Frequency Gap (MHz)          & 14                & 14 \\
		Fundamental Power (dBFS)     & -16.44            & -13.32 \\
		\hline
		IMD3 (dBc)                   & \textbf{-69.13}   & \textbf{-71.98} \\
        IMD2 / Multi-tone SFDR (dBc) &	-59.33           & -60.94\\
		\hline
	\end{tabular}
\end{table}

\subsection{Cross-talk between adjacent ADC channels}

\begin{figure}
\includegraphics[width=\columnwidth] {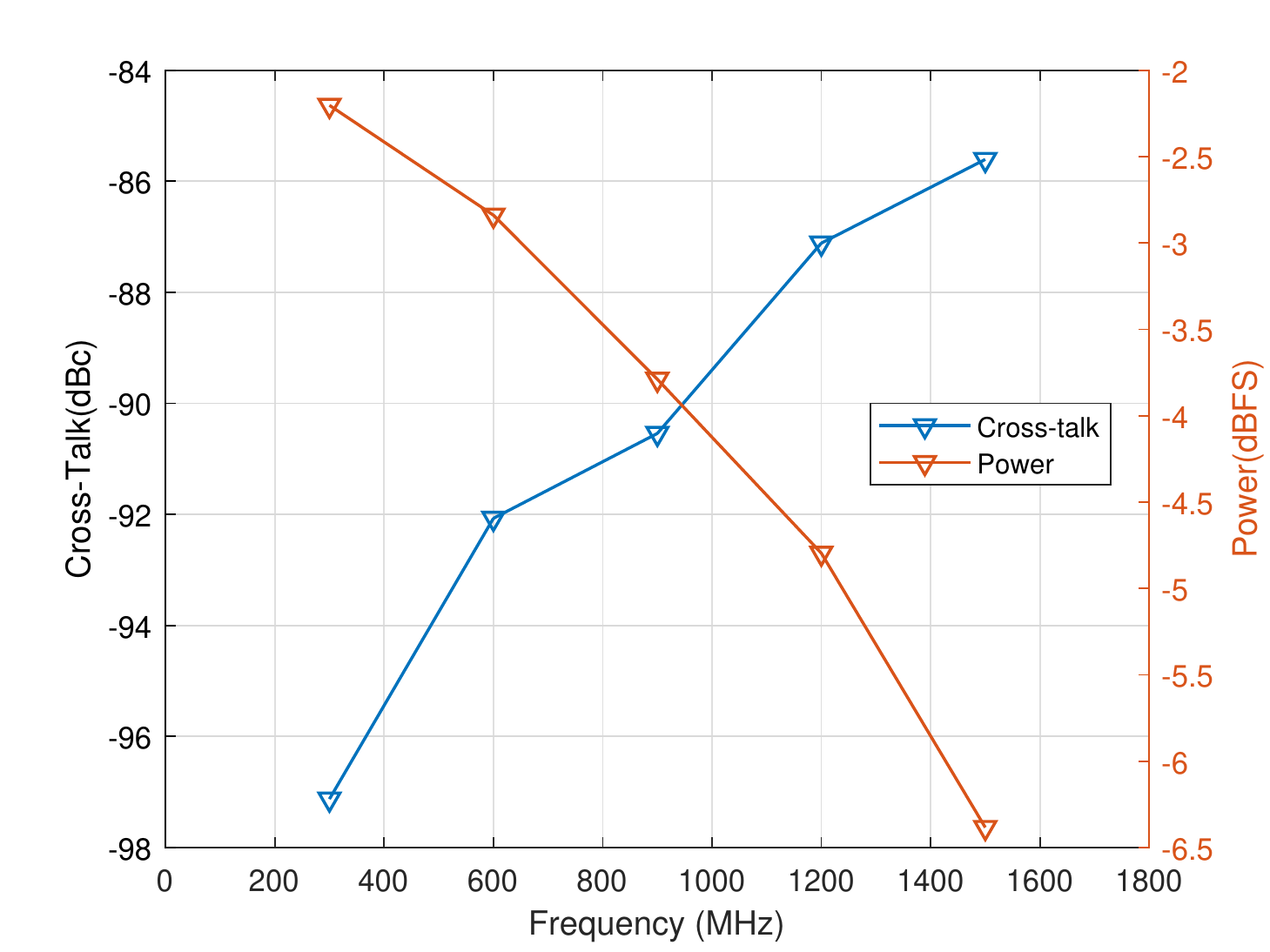}
\caption{The cross-talk signal power captured by ADC23 (left axis) and signal power captured by ADC01 (right axis).}
\label{aba:fig7}
\end{figure}

Radio astronomy instruments will likely use all of the eight ADCs integrated in the RFSoC. Cross-talk between channels will therefore be a concern if multiple channels are running simultaneously. In this test, the cross-talk between two adjacent channels, ADC01 and ADC23 in tile 224, was evaluated by injecting a signal in to ADC01 and measuring the signal coupled to ADC23, which had the input terminated. The firmware and software for the RFSoC were modified to operate both ADC channels at 4.096 GSPS and capture samples from both of the channels. The signal injected to the ADC01 used the same signal generator and frequencies as in Section \ref{sec:SFDR}, with a power level of at +5 dBm at the signal generator. In Figure \ref{aba:fig7}, the cross-talk captured by ADC23 is shown on the left axis and the signal power captured by ADC01 is shown on the right axis. The cross-talk level increases with frequency, even though less power was injected to ADC23 at higher frequency. The highest cross-talk is $-86$~dBFS at 1.5~GHz, which is lower still lower than the $-70$~dB cross-talk isolation specified in the Xilinx data sheet. Although the cross-talk measured here is lower than the specification, cross-talk might be worse with a different RF circuit design and routing of traces on the board. Therefore, the cross-talk level should be confirmed for any RFSoC-based board it is proposed to use, especially at high end of the frequency range. However, these tests show that cross-talk intrinsic to the RFSoC itself, even for ADCs on the same tile, is negligible.

\section{RFSoC-based spectrometer and performance evaluation}\label{sec:spectrometer}

\subsection{Spectrometer architecture}

All the tests in Section \ref{sec:ADCperformance} were performed by capturing ADC samples and processing the data obtained off-line. Since the ADC samples as fast as 4.096 Gsps, the capture can only be microseconds or milliseconds in length because of the bottleneck in memory and data throughput of the data links. For most radio astronomy receivers, the data need to be channelized and integrated or correlated in real time.  Therefore, it is essential to implement a spectrometer that operates in real-time for system performance evaluation and to be used as the basis for more complicated system development.

\begin{figure}
\includegraphics[width=\columnwidth] {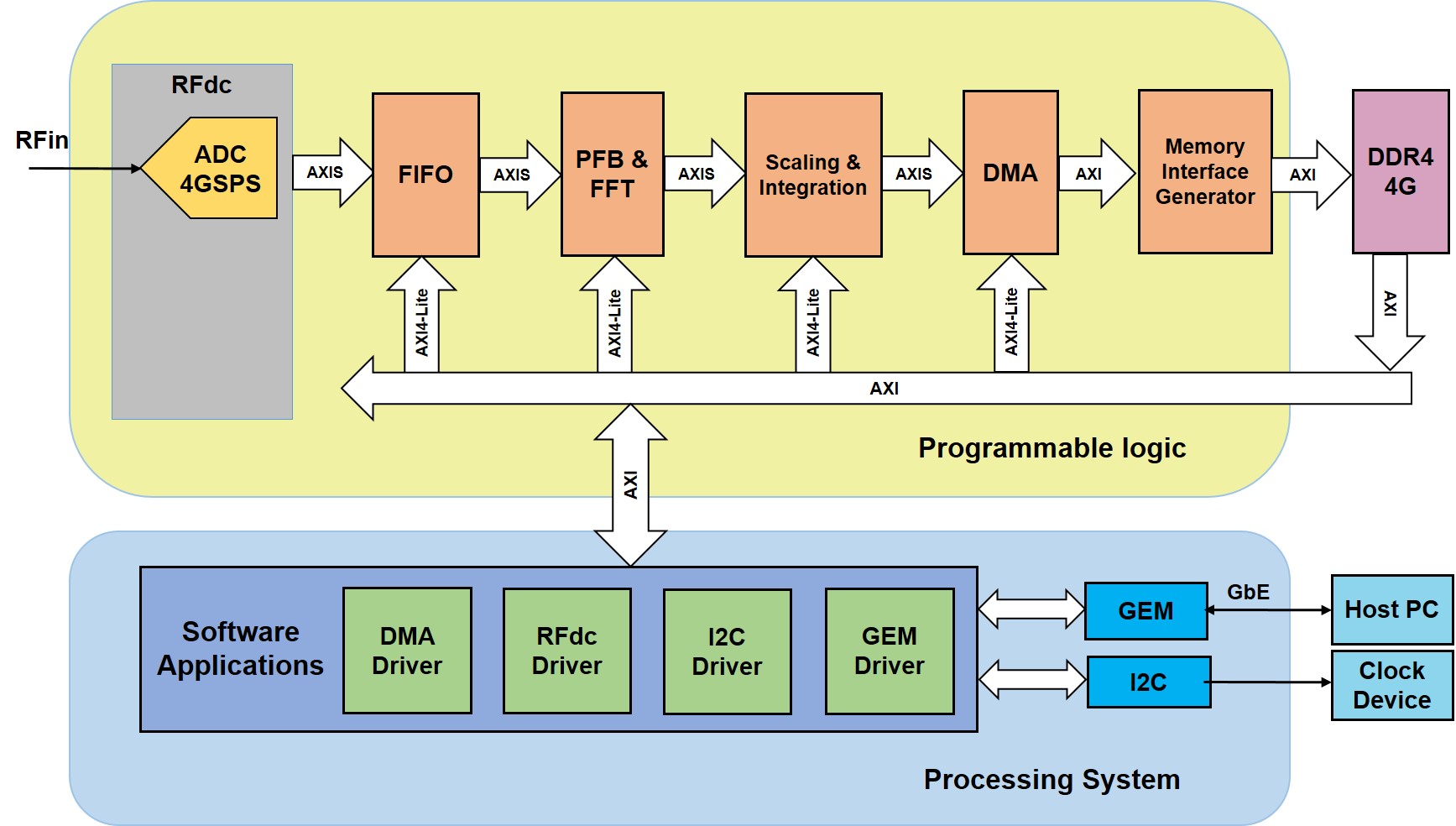}
\caption{Architecture of the FPGA spectrometer.}
\label{aba:fig8}
\end{figure}

We have implemented a single-channel spectrometer operating in real time with the architecture shown in Figure \ref{aba:fig8}. Compared with the ADC sample capturing firmware described in Section \ref{sec:setup}, additional digital signal processing (DSP) IP blocks have been inserted in the data path. The DSP IPs include a polyphase filter bank (PFB) with 3 taps and a 12-stage FFT, and scaling and integration IP blocks. The core of the PFB and FFT IP is generated using the System Generator-based toolflow from the Collaboration for Astronomy Signal Processing and Electronics Research (CASPER) \citep{2016JAI.....541001H}. The PFB has an FFT length of 4096, and the ADC samples at 4.096 GSPS, so the output spectrum ranges from DC to 2.048 GHz with a frequency resolution of 1 MHz. The scaling and integration IP blocks are custom-HDL based and have a configurable integration time and scaling factor through their software registers. Similarly to the ADC sample capturing system, the software application sets up and checks the operation of the IPs in programmable logic, and manages the data movement actions. In the final stage, the integrated and scaled spectrum is packetized and send to the host PC as UDP packets. On the host PC, the UDP packets of the spectrum are received and interpreted by a Matlab script, which can display the spectrum in real-time and save the spectrum integrated over any given time interval. 

\subsection{SFDR evaluation with the spectrometer }  

\begin{figure}
\includegraphics[width=\columnwidth]  {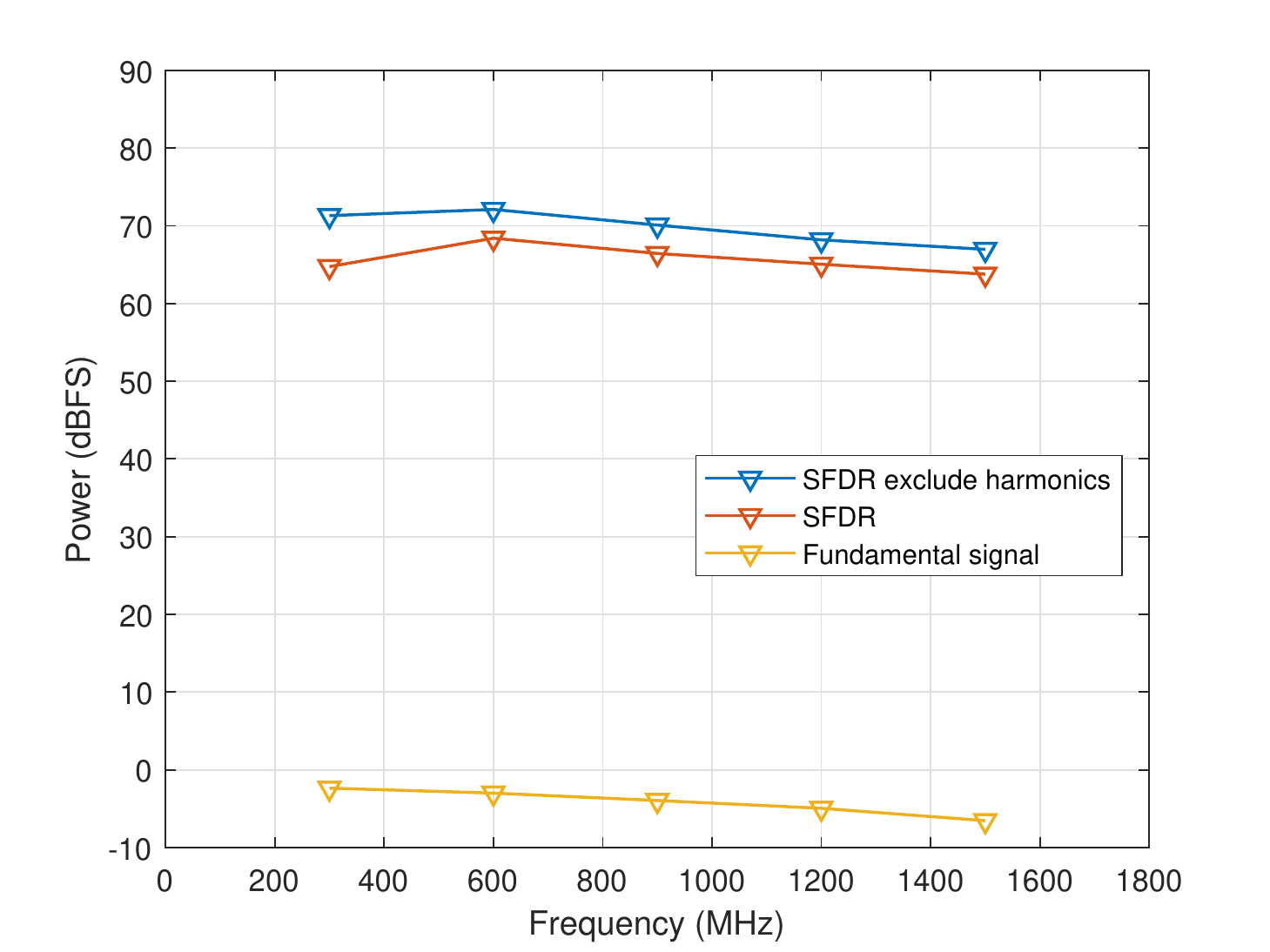}
\caption{SFDR measured by the spectrometer implemented in the FPGA. }
\label{aba:fig9}
\end{figure}

All the computation performed in Section \ref{sec:ADCperformance} used floating-point arithmetic in Matlab. However, when the algorithms are implemented in the FPGA, the operations are performed in fixed-point precision, which has the potential to limit the dynamic range due to truncation and rounding effects. To evaluate the performance of the spectrometer implemented in the FPGA, the SFDR test was performed again using the same tones as in Section \ref{sec:SFDR}, generated by the signal generator. The integration time was configured to be 1.024 ms, i.e. the sum of 1024 individual spectra. The output spectrum of the spectrometer was scaled to have the same fundamental signal power levels as in Section \ref{sec:SFDR}.  As Figure \ref{aba:fig9} shows, the SFDR and SFDR excluding the harmonics of the fundamental are about 10 dB lower than the SFDR results obtained in Section \ref{sec:SFDR}. The reasons for this performance loss require further investigation. It could be partially caused by the precision loss from floating-point to fixed-point implementations of FFTs, or from the difference in the FFT algorithm implementation between CASPER and Matlab. However, a SFDR of around 70 dB will be still adequate for the targeted radio astronomy applications.

\subsection{Noise level and stability evaluation with the spectrometer}

\begin{figure}
\includegraphics[width=\columnwidth]  {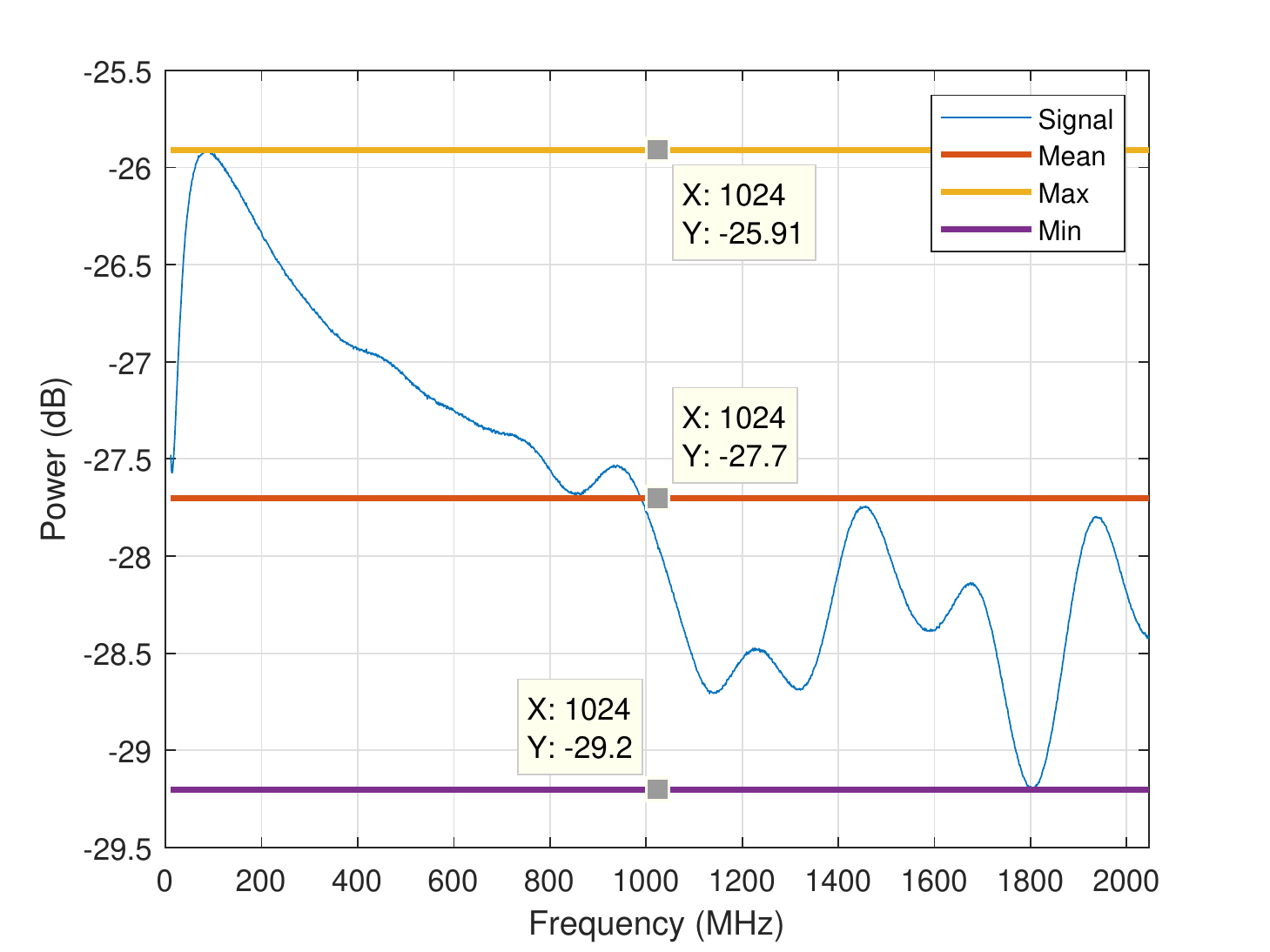}
\caption{Spectrum of the broadband noise source obtained by the FPGA spectrometer. }
\label{aba:fig10}
\end{figure}

The stability and noise level of the calculated spectrum are important parameters to characterise the high-level performance of the ADCs and spectrometer. An RF noise source (Noisecom NC1128B) with a frequency range from 10~MHz to 10~GHz and output power of 0~dBm was connected to the ADC input to evaluate the stability, flatness, and noise levels of the spectrometer. No anti-aliasing filter was used, in order to take measurements across the whole available bandwidth. The noise source has a flatness specification of $\pm$3 dB over the entire frequency range. The integration time was set to 1.024 s for this test, so more than 4 $\times$ $10^9$ samples were used to calculate one frame of the spectrum. Figure \ref{aba:fig10} shows the spectrum obtained from the noise-source signal. The flatness demonstrated is within  $\pm 1.8$~dB, better than the specification of the noise source, with no evidence of any artefacts introduced by the sampling or the spectrometer. There are no visible interleaving spurs at multiples of 512~MHz.

\begin{figure}
\includegraphics[width=\columnwidth]  {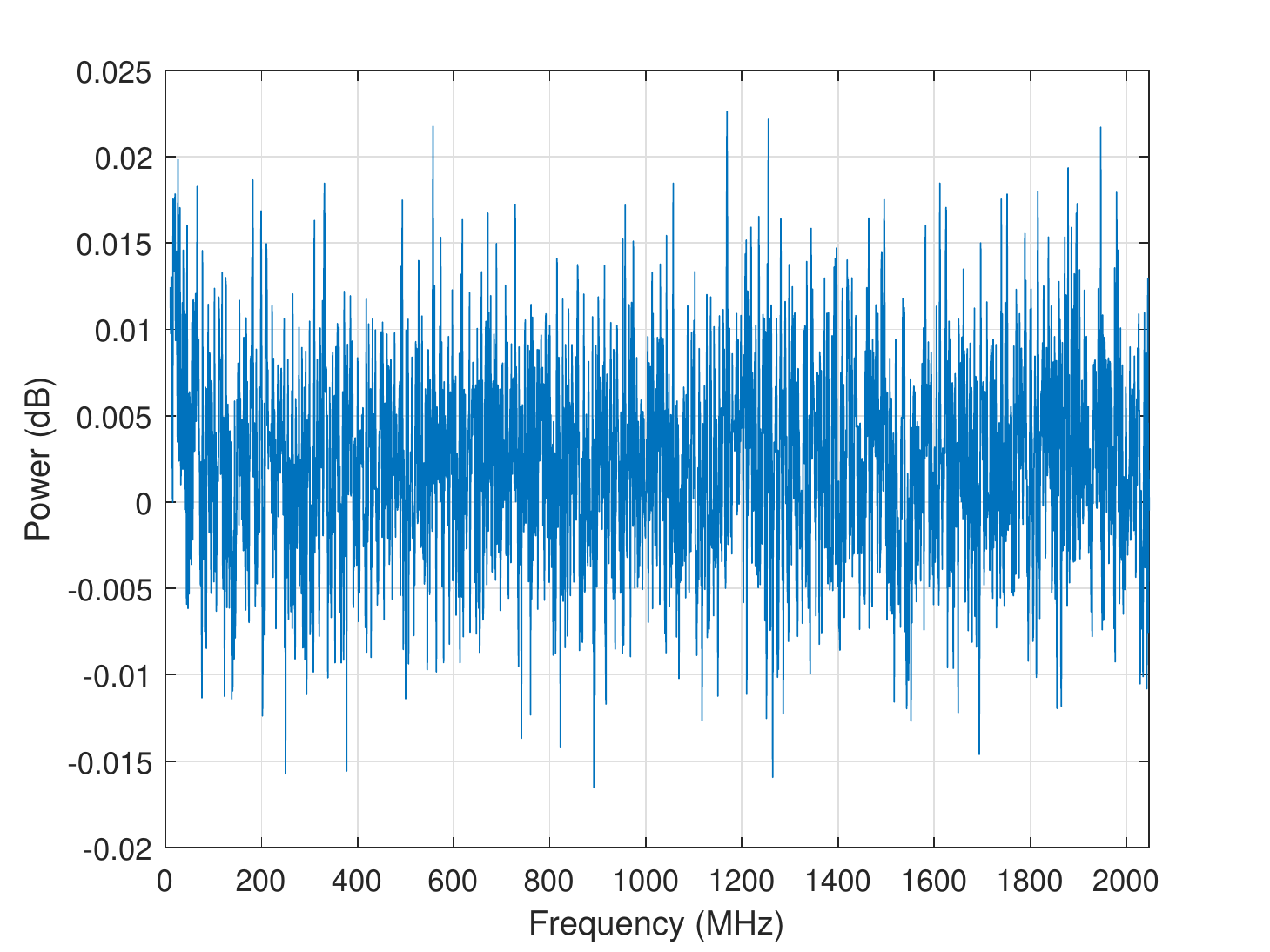}
\caption{The difference between two independent spectra of the broadband noise source taken 10 hours apart.}
\label{aba:fig11}
\end{figure}

In an actual instrument, the spectral flatness could be calibrated provided the stability of the spectrometer is good between calibration observations. To evaluate the stability of the spectrometer, two independent spectra were captured 10 hours apart, and subtracted from each other. The residual shown in Figure \ref{aba:fig11} only varies between $\pm$  0.025 dB, which implies the spectrometer and ADC operate at high stability and consistency. This would allow good calibration of spectral flatness.

\begin{figure}
\includegraphics[width=\columnwidth]  {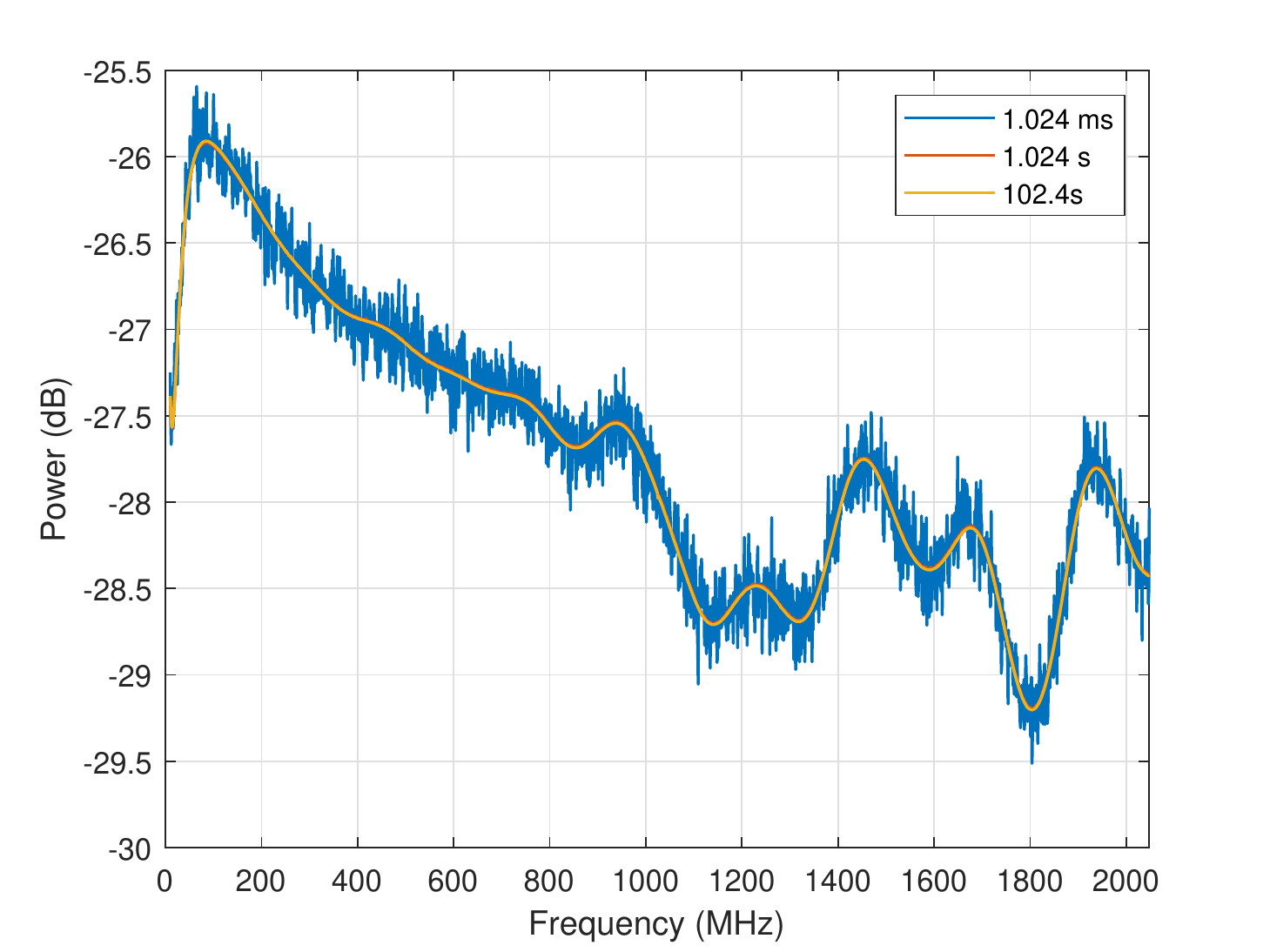}
\caption{Spectra of the noise source at different integration times. }
\label{aba:fig12}
\end{figure}

The integration time lengths of the target applications for the spectrometer we are developing can vary from milliseconds to hundreds or thousands of seconds. Therefore, the noise level of the spectrometer has been investigated at 1.024 ms, 1.024 s and 102.4s integration time lengths. The captured spectra, scaled by the integration time are shown in Figure \ref{aba:fig12}. All three spectra are within the range of flatness specification of the noise sources and none of them show any significant interleaving spurs or any other spurs. For the longest integration time of 102.4~s, over 4 $\times$ $10^{11}$ consecutive ADC samples, and $10^5$ individual spectra were used to calculate the spectrum shown in the figure. As shown in Figure \ref{aba:fig12}, there are no significant spurs at any of the frequencies in the spectra of the three integration time lengths. 

\begin{figure}
\includegraphics[width=\columnwidth]  {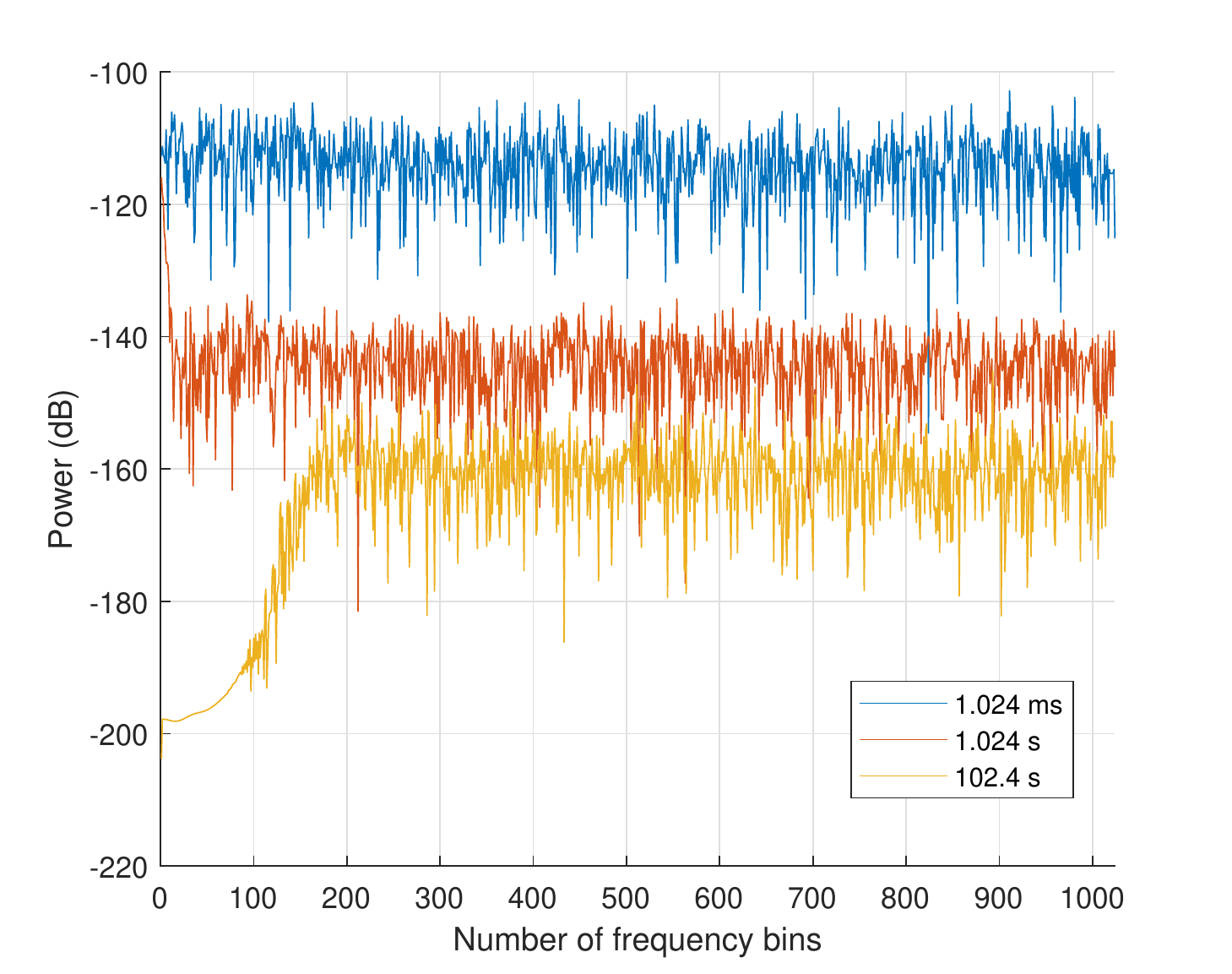}
\caption{Power spectra of the noise-source spectral measurements at different integration times after subtraction of the smoothed average spectrum.}
\label{aba:fig13}
\end{figure}

As Figure \ref{aba:fig12} shows, the spectra are dominated by the pass-band signal and become smoother as the integration time grows. To have a clearer view on the effect on the noise level of increasing the integration time, we have taken the power spectrum of the spectral traces for each of the three integration times. Before taking the power spectra, the pass-band signal profile has been subtracted from the spectra, so the noise signal can be characterized independently. The pass-band signal was determined by applying a $12^{\rm th}$ order Savitsky-Golay filter with frame length of 61 to the spectrum captured at integration time of 102.4 s. This filter applies a polynomial fit to short frames of data, removing small-scale noise but preserving the smooth features of the spectrum. As Figure \ref{aba:fig13} shows, the power spectra of the residual signals appear as white noise with decreasing level as the integration time increases. In Table \ref{aba:tbl2}, the mean values of the power spectra in Figure \ref{aba:fig13} are listed. For a spectrometer with high stability, the noise power level should reduce linearly with increasing integration time. From 1.024 ms to 1.024 s, integration time is increased by 1000 times, which should result in a 30 dB decrease in  noise level. We find that the mean value of the power spectrum is reduced by 30.2~dB. From 1.024 s to 102.4 s integration time, the mean power decreases by 20.1~dB, which is extremely close to the 20 dB theoretical value.

\begin{table}
	\centering
	\caption{The mean noise powers for spectra at different integration times.}
	\label{aba:tbl2}
	\begin{tabular}{lccr} 
		\hline
	    Integration time   &	Mean Value (dB)\\
		\hline
	    1.024 ms & -114.5\\
        1.024 s	& -144.7 \\
        102.4 s & -164.8 \\
		\hline
	\end{tabular}
\end{table}

\subsection{Spectral leakage of the spectrometer}

\begin{figure}
\includegraphics[width=\columnwidth] {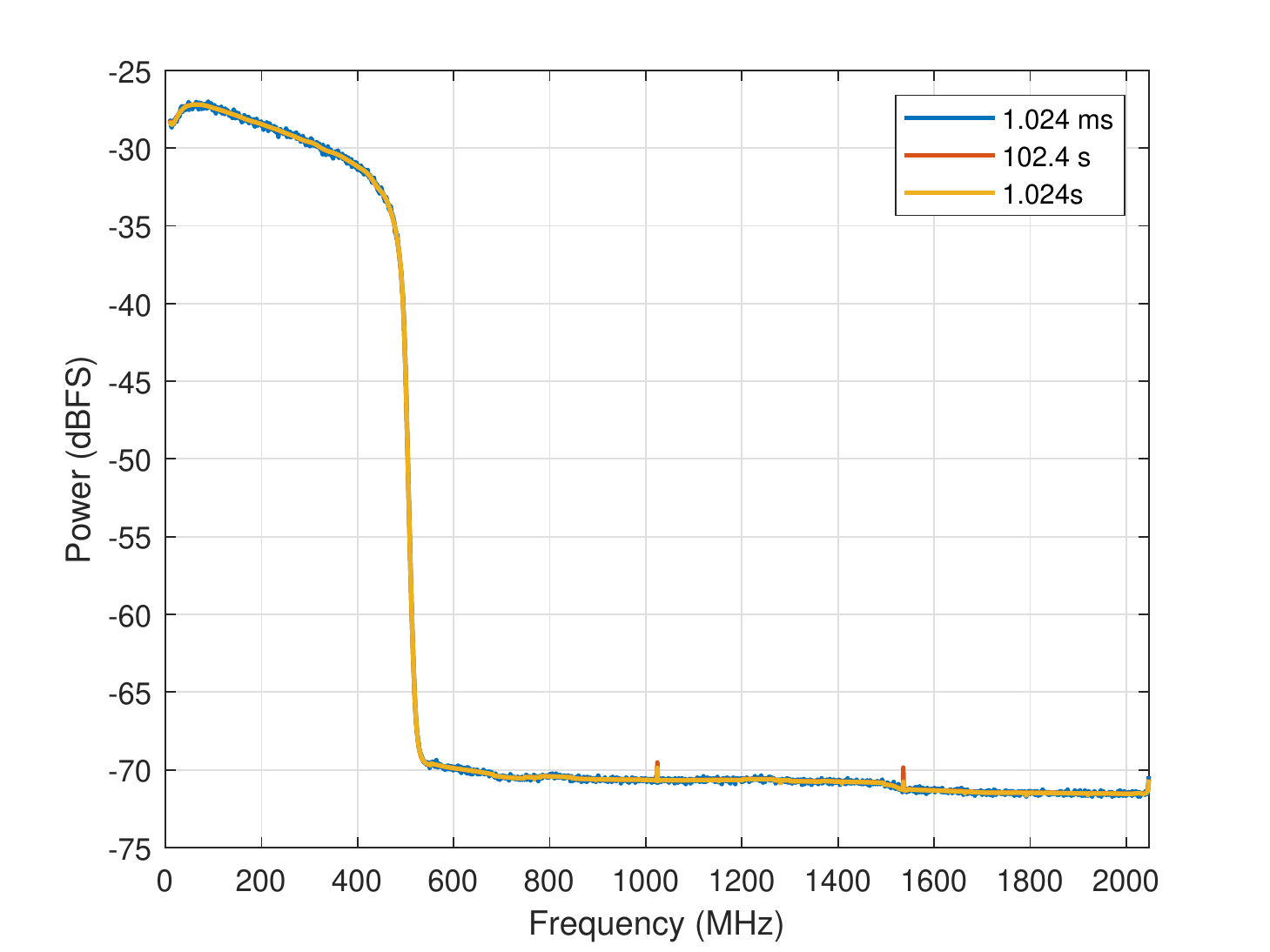}
\caption{Spectra of the noise source plus 500~MHz low-pass filter at different integration times.  }
\label{aba:fig15}
\end{figure}

Spectral leakage between the frequency bins is always a concern for a spectrometer or channelizer for a radio astronomy receiver. With the NC1128B noise source connected though a 500~MHz cutoff LPF, the leakage of the spectrometer was examined by looking at the power level outside the pass-band. A custom designed LPF with stop-band rejection of over 70~dB was used. The spectra of the noise source with the LPF at three different integration levels are shown in Figure \ref{aba:fig15}. The spectra are scaled to the full scale of the spectrometer. As Figure \ref{aba:fig15} shows, the spectrometer captured a 40~dB difference between pass-band and stop-band, and a clear cut-off edge between the pass-band and the stop band. The power level in the stop-band is around $-70$~dBFS, which is not as low as the $-90$~dBFS level shown in Section \ref{sec:noise_floor}. Simulations of the PFB using the {\tt pfb\_introduction} software package \citep{price2018spectrometers} show that this dynamic range limitation is due to the relatively low number of taps (three) used for PFB, and represents the accumulated leakage from power in the pass band to the stop band due to the sidelobes of the individual channel responses. The interleaved spurs at 1024 MHz and 1536 MHz grow as the integration time increases. However, even in the longest integration time of 102.4 s, the strongest spur in the stop-band is still as low as $-70$~dBFS.     

\subsection{Resource utilization of the spectrometer}

The utilization of the key FPGA resources shown in Table \ref{aba:tbl3} is around 10$\%$ for the single-channel spectrometer implementation. If all of the eight ADCs were used, some resource sharing and data-path optimization can be applied to reduce the overall utilization of flip-flops and BRAM, but it is difficult to significantly reduce the utilization of DSP slices. The DSP utilization with all the ADCs in use will be around 80$\%$. Therefore the frequency resolution that could be achieved by a RFSoC-based spectrometer using all the integrated ADCs will not be much higher than the 1 MHz achieved by the spectrometer implemented here. However this would be adequate for many applications, and packetized partly channelized data can always be shipped off the board for further processing using the high-speed data interfaces. 

\begin{table}
	\centering
	\caption{Resource utilization of spectrometer implementation}
	\label{aba:tbl3}
	\begin{tabular}{lccr} 
		\hline
	    Resource   &	Utilization	&   Available  &	Utilization ($\%$)\\
		\hline
	    Flip-Flop  &	99883       &	850560     &	11.74 \\
        BRAM       &	115	        &   1080       &	10.65 \\
        DSP	       &    399	        &   4272       &	9.34 \\
		\hline
	\end{tabular}
\end{table}

\section{Conclusions}\label{sec:conclusions}

We have characterized the ADCs integrated in the Xilinx ZU28DR RFSoC in the context of its use as a radio astronomy receiver. We have eliminated any concerns about spurs due to the interleaved design of the ADC, showing that these are removed by the internal calibration of the ADC timing. The experimental results also demonstrate the high dynamic range, spectral flatness, long-term stability and low crosstalk of the ADCs over their entire bandwidth. As well as off-line analysis of the data using floating-point arithmetic, we have demonstrated a real-time integer arithmetic spectrometer implemented on the RFSoC programmable logic. The performance of the ADC observed is more than adequate for our planned applications for single-dish spectral polarimeter-radiometers, and the FPGA resources available will allow us to implement 1~MHz channel resolution across the available bandwidth.  The performance will also be sufficient for many other radio astronomy applications, both for single-dish receivers and for the data acquisition and antenna-based signal processing in interferometers.  With the integrated RF ADCs, an RFSoC-based hardware platform can significantly shorten the design cycle and reduce the hardware cost of the digital system of a radio telescope compared to using discrete ADCs and FPGAs. RFSoC-based platforms are likely to be a increasingly used in radio astronomy digital backend development. 

\section{Data availability}

The data underlying this article will be shared on reasonable request to the corresponding author.



\bibliographystyle{mnras}
\bibliography{MNRAS_RFSoC} 





\bsp	
\label{lastpage}
\end{document}